\documentclass{aa}  

\usepackage[T1]{fontenc}
\usepackage{graphicx}   % Including figure files
\usepackage{amsmath}    % Advanced maths commands
\usepackage{amssymb}    % Extra maths symbols
\usepackage{natbib}
\usepackage{multirow}
\usepackage{ulem}

\usepackage{stackengine}

\newcommand\xrowht[2][0]{\addstackgap[.5\dimexpr#2\relax]{\vphantom{#1}}}

\usepackage{xcolor}

\newcommand{\mbh}{$M_{\mathrm{BH}}$}
\newcommand{\dotm}{$\dot{m}$}

\newcommand{\incl}{$i$}

\newcommand{\Hb}{H{\small $\beta$}}
\newcommand{\Ha}{H{\small $\alpha$}}
\newcommand{\MgII}{\ion{Mg}{II}}
\newcommand{\CIV}{\ion{C}{IV}}
\newcommand{\SiIV}{\ion{Si}{IV}}
\newcommand{\FeII}{\ion{Fe}{II}}

\newcommand{\HI}{\ion{H}{I}}

\newcommand{\Msun}{${\rm M_{\odot}}$}

\newcommand{\ergs}{erg\,s$^{-1}$}
\newcommand{\kms}{km\,s$^{-1}$}
\newcommand{\mergs}{\mathrm{erg\,s}^{-1}}
\newcommand{\mkms}{\mathrm{km\,s}^{-1}}
\newcommand{\Msyr}{${\rm M_{\odot}}$\,yr$^{-1}$}

\graphicspath{{./}{figures/}}

\begin{document}

\title{Deep absorption in SDSS J110511.15+530806.5}

% \correspondingauthor{Marcin Marculewicz}
% \email{marcin.marculewicz@gmail.com}
   \author{Marcin Marculewicz\inst{1}
        \and Marek Nikolajuk\inst{2}
        \and Agata Różańska\inst{3}
          }

\institute {Department of Astronomy, Xiamen University, Xiamen, Fujian 361005, China\\
\email{marcin.marculewicz21@xmu.edu.cn, marcin.marculewicz@gmail.com}
\and 
University of Bia\l{}ystok, Faculty of Physics, Cio\l{}kowskiego 1L Street, 15-245 Bia\l{}ystok, Poland
\and Nicolaus Copernicus Astronomical Center, Polish Academy of Sciences, Bartycka 18, PL-00-716 Warsaw, Poland
    }

   \date{Received XX, 2022; accepted XX, 2022}

\abstract
{}
{We study the origin of the anomalous deep absorption in a spectrum of the SDSS J110511.15+530806.5 distant quasar (z=1.929) obtained by the Sloan Digital Sky Survey in Data Release 7 of the optical catalog. We aim to estimate the velocity of absorbing material, and we show that this material considerably affects our measurements of the black hole (BH) mass in massive quasars with the use of common virial mass estimators.}
% methods heading (mandatory)
{The spectral shape of the quasar was modeled assuming that the 
accretion disk emission is influenced by a hot corona, warm skin, and absorbing material located close to the nucleus. The whole analysis was undertaken with XSPEC models and tools. The overall spectral shape was represented with the \texttt{AGNSED} model, while the deep absorption is well described by two Gaussians.}
% results heading (mandatory)
{The observed spectrum and the fitting procedure allowed us to estimate the BH mass in the quasar as $3.52 \pm 0.01 \times 10^9$~\Msun, the nonzero BH spin is $a_* = 0.32 \pm 0.04$, and the accretion rate is $\dot m=0.274 \pm 0.001$. The velocities of the detected absorbers lie
in the range of 6330-108135~\kms. When we consider that absorption is caused by the \CIV\ ion, one absorber is folding toward the nucleus with a velocity of 73887~\kms. We derived a 
BI index of about 20300~\kms \ and a mass outflow rate up to 38.5\% of the source accretion rate.}
% conclusions heading (optional), leave it empty if necessary 
{The high absorption observed in SDSS J110511.15+530806.5 is evidence of fast winds that place the source in the group of objects on the border with UFO (ultra-fast outflows), strong broad absorption line (BAL), and fast failed radiatively accelerated dusty outflow (FRADO). This absorption affects the BH mass measurement by two orders of magnitude as compared to virial mass estimation.}


\maketitle

%%%%%%%%%%%%%%%%%%%%%%%%%%%%%%%%%%%%%%%%%%%%%%%%%%%%%%%%%%%%%%%%%%%
%%%%%%%%%%%%%%%%% BODY OF PAPER %%%%%%%%%%%%%%%%%%%%%%%%%%%%%%%%%%%
\section{Introduction}
\label{sec:1}

Quasars (QSOs) are the brightest active galactic nuclei (AGN) and contain a supermassive black hole (SMBH) in their centers. They are powered by accreting matter that feeds the central engine. It is widely accepted that the mass of the SMBH in quasars is $> 10^6$ \Msun
\citep[e.g.,][]{Greene_Ho_contamination04}. It is important to know the BH masses when quasar phenomena are to be understood. The most reliable technique is reverberation mapping (hereafter RM). It is based on the study of the gas dynamics that surrounds the BH \citep[e.g.,][]{Blandford82, Peterson93, Peterson04RM, Bentz18, Shen19}. Most often, the gas located in the line-emitting region, such as the broad line region (BLR), is used in this technique because strong lines that are broadened by the Doppler effect are observed in the optical/UV spectra of many quasars. By measuring the time lags between emission in the continuum and in observed lines, we can directly estimate the size of the BLR and correlate this size $R$ with the quasar luminosity $L$. Furthermore, assuming that the line velocity corresponds to the virial velocity of the BLR, the virial mass of the central SMBH can be derived taking our ignorance of the structure and geometry of the line-emitting clouds into account. 

The modification of this method is the single-epoch (SE) virial BH mass estimator \citep[see][for a review]{Shen13}, which uses the $R-L$ relation previously 
that was determined through RM measurements. Assuming that this relation is universal, we can 
estimate the BLR size at any time for a source whose luminosity is known. This parameter combination with calibration coefficients allows us to estimate the BH mass using only the line measurement. 
Thus, these methods are powerful and frequently used because their approximation works very well, but they are expensive in terms of the number of high-resolution spectroscopic data needed, and are used on a case-by-case basis (and for only a few objects). However, the future development of the Black Hole Mapper\footnote{Black Hole Mapper: {\tt\string www.sdss.org/dr14/future/bhm/}} project will increase our knowledge and will help us to constrain the SMBH mass range well. 

An alternative less expensive but nondynamic method is the continuum spectral fit method. It is based on the well-grounded model of continuum emission from an accretion disk that surrounds an SMBH  \citep[e.g.,][]{SS73, NT73}. The nondynamic term here is used in the context that we do not need any gas velocity measurements. Only time-averaged continuum spectra are used and fit with the accretion disk emission model. One of the most important parameters in this model is the mass of the BH. \cite{Marculewicz2020} showed that the continuum-fit method works well for calculations of the BH masses of weak emission-line quasars (WLQ). We have shown that the masses of weak emission-line quasars that were determined on the bases of the full width at half maximum (FWHM) of the \Hb\ line are underestimated. According to our analysis, both methods, the RM and the SE, give BH masses that are underestimated by four to five times when FWHM is lower than $10^4$ \kms\
in these types of objects. In the above paper, we proposed an alternative equation to provide a more precise constraint of the BH masses in these types of objects. The continuum emission model we used for our analysis was composed of an accretion disk component together with starlight and emission from a dusty torus. 

In this paper, we apply the continuum-fit method of BH mass estimation to the peculiar quasar SDSS J110511.15+530806.5 (hereafter J1105), which exhibits deep absorption in the optical/UV domain. 
The source was first identified by Sloan Digital Sky Survey (SDSS) Data Release 6 obtained on September 17, 2007. The BH mass in this quasar reported by \citet{Shen2011qsocatalog} is $\log M_{\rm BH} = 11.205$  solar masses, indicating that this object is a behemoth with an enormous mass. Notwithstanding, \cite{Shen2011qsocatalog} indicated that the mass may be erroneous, and these virial masses for individual objects have to be interpreted with caution \citep{Shen_2008MBHbiases}. Considering all of this and taking the conclusions of our previous paper \citep{Marculewicz2020} into account, we used a phenomenological model of disk emission with a hot corona and warm skin affected by absorbing gas to fit the continuum spectrum of J1105 that was obtained by the SDSS Data Release 7  \citep{SDSSDR7} to constrain the BH mass in this quasar. 

The second step in our analysis aims to explain the deep absorption that is visible in the optical/UV spectrum of J1105. We modeled this unusual absorption and discuss eventual scenarios of the origin of the absorbing material in 
this peculiar quasar. The paper is organized as follows: section \ref{sec:2} consists of the source description and a summary of our knowledge of the origin of absorption in QSOs. Section \ref{sec:3} presents the data reduction, model description, and analysis procedure we used to fit a model to the observations. 
Results are presented in Sec~\ref{sec:4}, and final conclusions are given in Sec~\ref{sec:5}. 
We computed the luminosity distances using the standard cosmological model with $H_{0}$ = 70 \kms\ Mpc$^{-1}$, $\Omega_{\Lambda}$ = 0.7, and $\Omega_{\rm M}$ =
0.3 \citep{Spergel}.

\section{SDSS J110511.15+530806.5 (J1105)}
\label{sec:2}

We searched the \cite{Shen2011qsocatalog} catalog for the sources with the most massive SMBHs. The catalog represents the properties of the 105,783 quasars in the SDSS Data Release 7 quasar catalog \citep{SDSSDR7}. It contains the continuum and emission line measurement of \Ha\,, \Hb\,, \MgII,\, and \CIV. We present recent results for J1105, which aroused our curiosity not only through its high BH mass, but also through peculiar spectral absorbing features in the optical/UV domain. J1105 has not been studied extensively. Any publicly available data were mainly taken from a wider sample of objects, and no particular attention was focused on the J1105 quasar. 
We are the first to study J1105 individually.
We searched for data of this quasar that were obtained at other energy ranges, but found none. Any possible X-ray detections by the ROSAT, INTEGRAL, or Swift missions show an offset of about 3 degrees, which is not acceptable as a point-source detection.
Nevertheless, we have tried to detect the source from the field of view of these satellites, but we failed because for each 
observation, source detection was  impossible at the significance level of a statistical detection above the huge background emission.
Therefore, we failed to obtain any limits for source emission in the X-ray band.

\subsection{Source description}
\label{source}

\begin{table}[ht]
\caption{J1105 parameters, position, redshift, and extinction.}
\begin{center}
\begin{tabular}{ c|c|c|c|c}
%\begin{tabular}{ c|c|c|c|c|c }
\hline %\extravspace
RA & Dec  & {$z_{\rm spec}$} & {$A_{\rm V}$} & E(B-V) \\
%& log{\mbh} & log \dotm  \\
(J2000.0) & (J2000.0) &  & [mag]  \\ 
%& [log \Msun] \\
\hline %\extravspace
166.296486$^{\circ}$ & 53.135186$^{\circ}$  & $1.929$ & 0.027 & 0.0087\\
%& $11.205$ & -4.713  \\
%& & $\pm0.0015$ & \\
%& $\pm 0.216$ & \\
\hline
\end{tabular}
\label{tab:J1105_para}

\end{center}
\end{table}

J1105 lies in the Ursa Major constellation at a comoving radial distance of $5075$ Mpc.
Its redshift based on the spectrum, $z_{\rm spec}$, is $1.929$ \citep{Meusinger2016}. The Galactic extinction measured in V and K filter toward J1105 is $A_{\rm V}$ = 0.027 mag and $A_{\rm K}$ = 0.003 mag \citep{Schlafly2011}.  

The observed spectrum of the quasar was corrected for Galactic reddening with an extinction law. This extinction curve is usually parameterized in terms of the amount of interstellar absorption at V color, $A_V$, and a measure of the relative extinction between B and V band, $R_V = A_V /E(B - V)$. The value of $R_V$ varies from 2.6 to 5.5 in the measurements of the diffuse interstellar medium \citep{Cardelli89,Fitzpatrick99}. The mean value is 3.1, and we used this to calculate $E(B-V)$. The $A_V$ value was taken from the NED\footnote{The NASA/IPAC Extragalactic Database (NED): {\tt\string ned.ipac.caltech.edu}} based on the dust map created by Schlegel et al. (1998). According to this, the color excess $E(B-V)$ toward J1105 quasar is 0.0087. We present the source position and reddening in Table~\ref{tab:J1105_para}. The measured mass of its SMBH is $\log M_{\rm BH} = 11.205$, and the Eddington accretion rate is  $\log \dot{m} = -4.713$ \citep{Shen2011qsocatalog}.

The monochromatic source luminosity was reported by \citet{Chen18_J1105_lum} to be  $\log \lambda L_{\lambda3000} = 45.534$ [\ergs]. To obtain the bolometric luminosity, the correction factor should be taken into account, as described by \citet{Netzer2019_bol_corr},
\begin{equation}
L_{\rm bol}=k_{\rm bol} \ \lambda L_{\lambda3000} = e \left[\frac{\lambda L_{\lambda3000}}{10^{42} \mergs}\right]^f \lambda L_{\lambda3000}.
\label{eq:bol_corr}
\end{equation}
When the Balmer continuum and the {\FeII} lines are included in the observed $\lambda L_{\lambda3000}$ , the obtained bolometric luminosity is  $\log L_{\rm bol}$ = 46.106 [\ergs] (i.e., $e = 19$, $f = -0.2$). Without information on whether the Balmer continuum and/or the {\FeII} lines were included in the luminosity at 3000\AA, we obtain $\log L_{\rm bol}$ = 46.225 [\ergs] (i.e., $e = 25$, $f = -0.2$).

J1105 was classified as a strongly reddened narrow-line quasar. \cite{Jiang_2013} argued that this unusual reddening law is based on the speculative assumption of an exotic dust grain size distribution that lacks large grains. \citet{Meusinger2016} studied J1105 and indicated that the reddening of the quasar is steeper than in the Small Magellanic Cloud, and perhaps even steeper than for the galaxy IRAS 14026+4341. The authors were also unable to determine a good reddening or extinction solution for J1105. 

\begin{figure}
  \centering
\includegraphics[width=0.50\textwidth]{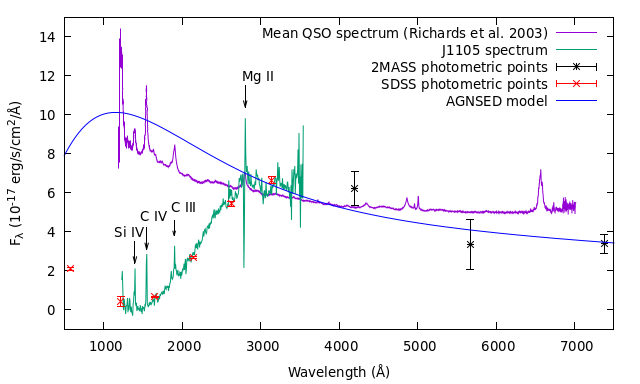}
 \caption{J1105 spectrum in the SDSS Data Release 7 (green line) in comparison with the mean QSO spectrum (purple line) by \cite{Richards_2003}. Both spectra were correlated regarding the \MgII\, line. In addition, we present earlier photometric points by the SDSS \citep{SDSSDR6} and 2MASS \citep{2006_Skrutskie} surveys, and the final broadband fit model is shown with the blue line.}
\label{fig:sdss_J1105_plus_richards}
\end{figure}

The spectrum of J1105 extracted by us from SDSS Data Release 7  (green line) in comparison with the mean QSO spectrum \citep[purple line;][]{Richards_2003} is presented in Fig.~\ref{fig:sdss_J1105_plus_richards}. The figure clearly shows the significant absorption of the J1105 spectrum, the nature of which is not yet known. Photometric data of the earlier SDSS Data Release 6 \citep{SDSSDR6} are shown as red points clearly follow this trend. In addition, the black points are photometric points from the 2MASS survey \citep{2006_Skrutskie}. We aim to explain this unusual absorption on the way from the source to the observer in this paper.

\subsection{Examination and evaluation of the absorption}

Absorption lines are visible in a significant number of QSO spectra. Quasars can be divided into BAL sources and narrow absorption line (NAL) sources based on their line widths. The NALs are associated with line velocities of just a few hundred \kms \citep[e.g.,][]{Wild_abs2008}; BAL QSOs have higher line velocities (a few thousand to even tens of thousand \kms\ \citep{Stone_abs_2019}. About 60\% of the QSO spectra exhibit the NAL features \citep[e.g.,][]{Vestergaard2003_CIV_EW}, whereas BALs are detected in only 20\% of the sources \citep[e.g.,][]{Knigge2008}. 

Strong quasar absorption line systems have two main origins:
(1) intervening material along the line of sight (e.g., galaxies), and (2) gas associated with the central engine of the quasar (e.g., winds from the accretion disk, galactic-scale outflow or inflow, nearby dwarf galaxies, and molecular clouds). Because of their properties, BAL QSOs appear to be associated with outflowing matter, whereas NAL systems represent intrinsic or intervening
absorbing material.

\cite{Stone_abs_2019} suggested the division of NALs with regard to the velocity of the absorbed matter. They distinguished between intervening absorption, regarding the number density per unit redshift ($dN/dz$), and intrinsic absorption, regarding the number density per unit velocity ($dN/d\beta$, $\beta = \frac{V_{\rm abs}}{c}$). However, the idea that an NAL is represented by intrinsic or intervening material is still debated.
Moreover, \cite{Stone_abs_2019} classified the quasar absorption line systems depending on the exact values of their velocities. 
Systems within $V_{\rm out} \lesssim$ 3000 \kms\ are called associated 
absorption lines (AALs). Systems with a velocity separation $3000 - 12000$ \kms\ are dominated by intrinsic material in the form of winds blowing from the QSO. Systems with a velocity $\gtrsim 12000$ \kms\ 
are disconnected from the quasar (see references therein).

The mean balnicity index (BI), which is the sum of the equivalent width of all contiguous BAL troughs across the observed velocities, was calculated for \CIV\ or \SiIV\ absorption doublets. It 
ranges from 1000 to 4000 \kms\ in more than 21000 BAL quasars \citep[SDSS catalogs; e.g.,][]{Gibson09,Allen2011,Paris2018}.
Higher values were reported for BAL quasars 
by \cite{Rankine2020}, showing that  BI = 6007 \kms\ for SDSS J140440.85+632705.9
and 9896 \kms\ for J105931.96+463638.9. The strongest absorbers indicate BI $\lesssim$ 23 000 \kms\ \citep{Paris2018}. 

The purpose of this work is to describe the exceptional nature of the high-mass quasar J1105. The high absorption that has not been observed before caught our attention. In the section below, we specify the absorbing gas velocity and discuss whether the observed absorption is associated with or disconnected from the quasar.
The velocity of the absorbing material is derived assuming 
that the absorber is a large-scale wind, and the shift is caused by the Doppler effect using the relativistic formula
\begin{equation}
   V_{\rm out} =  \beta \, c = \frac{(1+z)^2-1}{(1+z)^2+1} \, c,
   \label{eq:1}
\end{equation}
where  $V_{\rm out}$ is the speed along the line of sight, and $c$ is the speed of light. Furthermore, taking into account that the absorption is modeled by the Gaussian function, we were able to estimate the ion column densities, $N_j$, of the absorbers. In an optically thin regime, for a fixed value of the ionization parameter and metallicity inside the absorber
 \citep{Saez2021_NH}, we have
\begin{equation}
    N_j \approx \frac{m_{\rm e} c^2}{\pi e^2 (f_{\rm b} \lambda^2_{\rm b}+f_{\rm r} \lambda^2_{\rm r})} EW\ ,
    \label{eq:2}
\end{equation}
where b and r stand for parameters of the blue and red \CIV/\MgII\ doublet, respectively, $f$ is an oscillator strength taken from \cite{Griesmann2000Carbon} and 
\cite{Morton2003Magnesium}, and $EW$ is the equivalent width of the fitted absorbing Gaussian function. 

When the gas velocity is measured, the interaction of the outflow with the surrounding material can be estimated using the commonly known formula 
\begin{equation}
\dot M_{\rm out} \simeq \frac{\alpha(r) L_{\rm bol}}{V_{\rm out} c},
\label{eq:mass_outflow}
\end{equation}
where $\dot M_{\rm out}$ is the mass outflow rate, and $\alpha(r)$ is the radial dependence of the optical thickness of the flow. In the simplest example, the flow is Compton thick at all radii,
hence $\alpha(r)=1$. We discuss all these physical parameters in the context of quasar J1105 in Sec.~\ref{sec:4}.

%=====================================================================
\section{Data and analysis method}
\label{sec:3}

We extracted the spectrum of J1105 at visible wavelengths from Data Release 7 of the SDSS optical catalog \citep{SDSSDR7} (see Fig. \ref{fig:sdss_J1105_plus_richards}). The observed spectrum of the object was corrected for Galactic reddening with the extinction law of \citet{Cardelli89,Fitzpatrick99}. The $A_{\rm V}$ value is taken from the NED database based on the dust map created by \citet{Schlegel98}. The spectrum of J1105 in comparison with the mean QSO spectrum shows extraordinary absorption in the range from $\sim 1200$ up to $\sim 2600$ \AA.

As a first step to explain the absorption, we used different extinction laws for Milky Way dust and grains. \cite{Czerny_extinction_2004} assumed different carbon grain and silicate dust temperatures in galaxies, and they showed that the different grain radii may explain various extinction curves for AGNs. Their study mainly considered amorphous carbon grains as an explanation for red quasars. On the other hand, \citet{Wickramasinghe1998} postulated that microdiamonds may exist in the interstellar medium, and thus the excess of the interstellar extinction at the UV can be observed. 
Unfortunately, none of the examined laws of extinction for the spectrum of J1105 studied by us worked for an explanation of the deep and broad absorption.

\begin{figure}[t]
  \centering
    \includegraphics[width=0.47\textwidth]{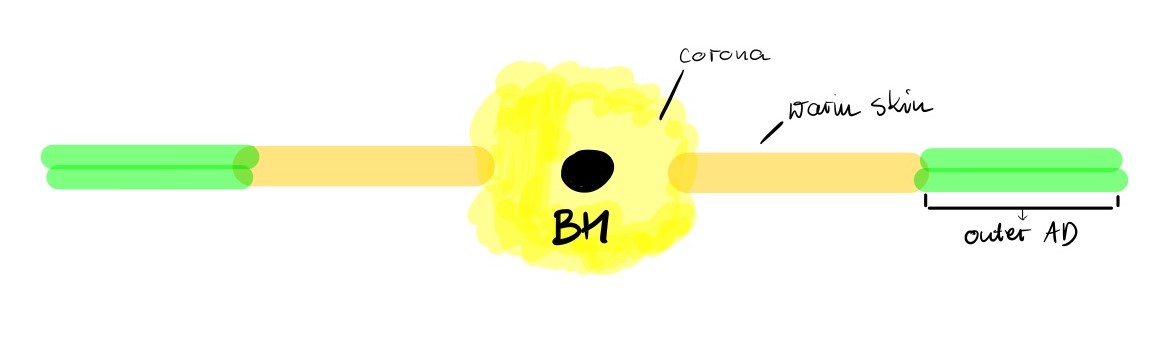}
 \caption{Vicinity of the BH with the geometrically thin accretion disk (green), 
 the warm skin (dark yellow), and the inner hot corona (yellow) system. 
 This radiatively stratified flow was considered to compute the intrinsic spectrum of 
 the QSO with \texttt{AGNSED} model used in our data-fitting procedure.}
 \label{fig:AD_corona_vicinity_general}
\end{figure}

As a second step, we decided to search the location of the absorbing gas, with the assumption that the quasar presents typical emission as an active galactic nucleus. To do this, the continuum spectrum of the source can be described with absorbed emission of a geometrically thin accretion disk with the inner hot corona and the warm skin. Because the emission from a hot corona and warm skin is observed in the X-ray domain, we find a phenomenological model of the continuum emission only in the XSPEC fitting package. Therefore, we decided to model the spectrum of J1105 and its global absorption using XSPEC tools v12.10.1. The XSPEC package is a command-driven, interactive, spectral-fitting program, designed to be completely detector independent so that it can be used for any spectrometer \citep{XSPEC1996}. 
For the active quasar engine, we used a standard geometry of the inner hot corona, warm skin, and the outer accretion disk, as presented in Fig. \ref{fig:AD_corona_vicinity_general}. 
This geometry is not yet fully confirmed because we do not know the physical mechanism that can separate the warm skin in the radial direction. A slab geometry of the warm skin and cold accretion disk in a vertical direction is often considered because it can be created in a magnetically supported disk,
as shown by \citet{2020-Gronkiewicz}, but the available spectral models to work with the data are still simpler and do not take vertical heating of the warm skin into account. 

One of the most plausible models available in XSPEC is the \texttt{AGNSED} model \citep{KubotaDone_2018}. It describes the emission from all regions presented in Fig. \ref{fig:AD_corona_vicinity_general} without any assumption about eventual gas heating. The modeled emitted spectrum can be adapted to the source parameters, and then eventually absorbed by the line-of-sight material. In the subsection below, we present the method and describe the model in full. 

\section{Modeling with XSPEC}
\label{sec:xspec}

From raw SDSS data, we reconstructed data for XSPEC compliance. We generated spectra for XSPEC and response matrices following standard procedures using \texttt{ftflx2xsp} to create a spectral (\texttt{PHA) file} and dummy response (\texttt{RSP) file}. This allowed us to model and fit J1105 using models available in XSPEC. For the purpose of fitting the model to the data, we binned the optical spectrum before transforming it into XSPEC, even though it did not influence the overall shape, and only emission lines were reduced to the level of the continuum. The binning procedure created bins between 30-100~\AA\ depending on the spectral window. The final spectrum ready for XSPEC analysis had 26 bins, which is sufficient for our research goal.

Our basic emission model was \texttt{AGNSED.} It is dedicated to reflect the intrinsic spectrum of QSOs. Here, the flow is assumed to be radially stratified, emitting as a standard disk blackbody from $R_{\rm out}$ to $R_{\rm warm}$, as warm Comptonization from $R_{\rm warm}$ to $R_{\rm hot}$, and then makes a transition to the hard X-ray emitting hot Comptonization component from $R_{\rm hot}$ to $R_{\rm ISCO}$ (innermost stable orbit radius). The outer accretion disk is optically thick and geometrically thin, as described by the Novikov-Thorne equations \citep{NT73}. For the warm Comptonizing region, the \texttt{AGNSED} model adopts the passive-disk scenario tested by \cite{Petrucci2018comptonization}. 
The inner hot corona is optically thin and emits energy in the hard Comptonized end of the spectrum.

The \texttt{AGNSED} model depends on many source parameters: BH mass, $M_{\rm BH}$, accretion rate $\dot m$ given in units of Eddington accretion rate, inclination angle $i$, dimensionless BH spin $a_*$, two coronal temperatures $kT_{\rm e, hot}$ and  $kT_{\rm e, warm}$, photon indices of the two Comptonized media $\Gamma_{\rm hot}$ and $\Gamma_{\rm warm}$, and on the geometrical parameters: radius of the hot corona $R_{\rm hot}$, 
radius of the warm skin $R_{\rm warm}$, outer accretion disk radius $R_{\rm out}$, and $H_{\rm tmax}$ , which is an upper limit of the scattering for the hot Comptonization component and is usually equal  to $R_{\rm hot}$. All these distances are given in units of gravitational radius $r_{\rm g}$. Furthermore, we fixed the distance $D$ in Mpc, the source redshift $z_{\rm spec}$, and fit the normalization of the overall model. In addition, the newest version of the model allows us to switch the parameter for the reprocessing. When the reprocess parameter is 0, the radiation coming from the hot corona is not reprocessed by the warm skin and the outer cold disk. When this parameter is 1, the hard X-ray flux from the hot corona is added to the local warm skin and cold disk flux, assuming a reflected albedo, and depending on the distance from the BH  through the purely geometrical formula 
\citep[see Eq. 5 in][]{KubotaDone_2018}. The seed photon temperature for the hot Comptonization was calculated internally. The \texttt{AGNSED} model does not take the color temperature correction into account because it does not calculate 
atmospheres. 

The intrinsic QSO spectrum is absorbed by many regions, including absorption in 
our Galaxy. We used the \texttt{redden} model for the reddening in the Milky Way, as described by \cite{Cardelli89} and given by the typical E(B-V) parameter. Furthermore, the photoelectric absorption in our Galaxy was taken into account with the \texttt{phabs} model with the equivalent hydrogen column density $N_{\rm H}$ as the only model parameter. 

After the two main absorptions in our Galaxy were taken into account, we were left with the possibility of adding additional clouds to study the distribution of the intrinsic absorption regions in J1105. We modeled each absorbing medium with a Gaussian function in order to find the relative velocity (i.e., redshift) of each region. Each Gaussian function is represented by three parameters: the energy $E$ in keV, the width $\sigma$ in keV, and the line depth $\tau$ , which is the optical depth in the line center.

To summarize, our total model fit to the data is \texttt{redden*phabs*AGNSED*gabs}(N), where N is the number of absorbing Gaussian functions required by our data. A close examination of the quasar spectrum indicated that at least two absorbers are required, that is, N=2. In this case, our total model has 23 parameters, and we decided to fix some of them to make our analysis plausible. We fixed E(B-V) for the J1105 quasar to 0.0087. Furthermore, we estimated $N_{\rm H}$ measured in the Milky Way \footnote{NASA's HEASARC N$_{\rm HI}$ column density calculator : \\
{\tt \string heasarc.gsfc.nasa.gov/cgi-bin/Tools/w3nh/w3nh.pl}} 
toward J1105 as $8.63 \times 10^{19}$~cm$^{-2}$. In addition, the outer radius of the accretion disk was fixed in our fitting procedure, while the radii of the hot corona and the warm skin were derived in the fitting procedure. Because we lack X-ray data of  J1105, we also fixed the temperatures of the  hot corona and warm skin and the hot corona photon index. The warm corona photon index stayed free as it can be influenced by fixed reddening and Galactic absorption. Finally, both the distance to the source and its redshift were fixed during the fitting procedure. We kept the reprocessing parameter equal to 1 and the albedo equal to 0.3 in our analysis. After fixing these parameters, we fit the data with 14 free parameters, which provides us with 12 degrees of freedom. 

%=====================================================================
\section{Results}
\label{sec:4}

\begin{figure}[ht]
  \centering
  \hspace*{-0.3cm}
   \includegraphics[width=0.6\textwidth]
   {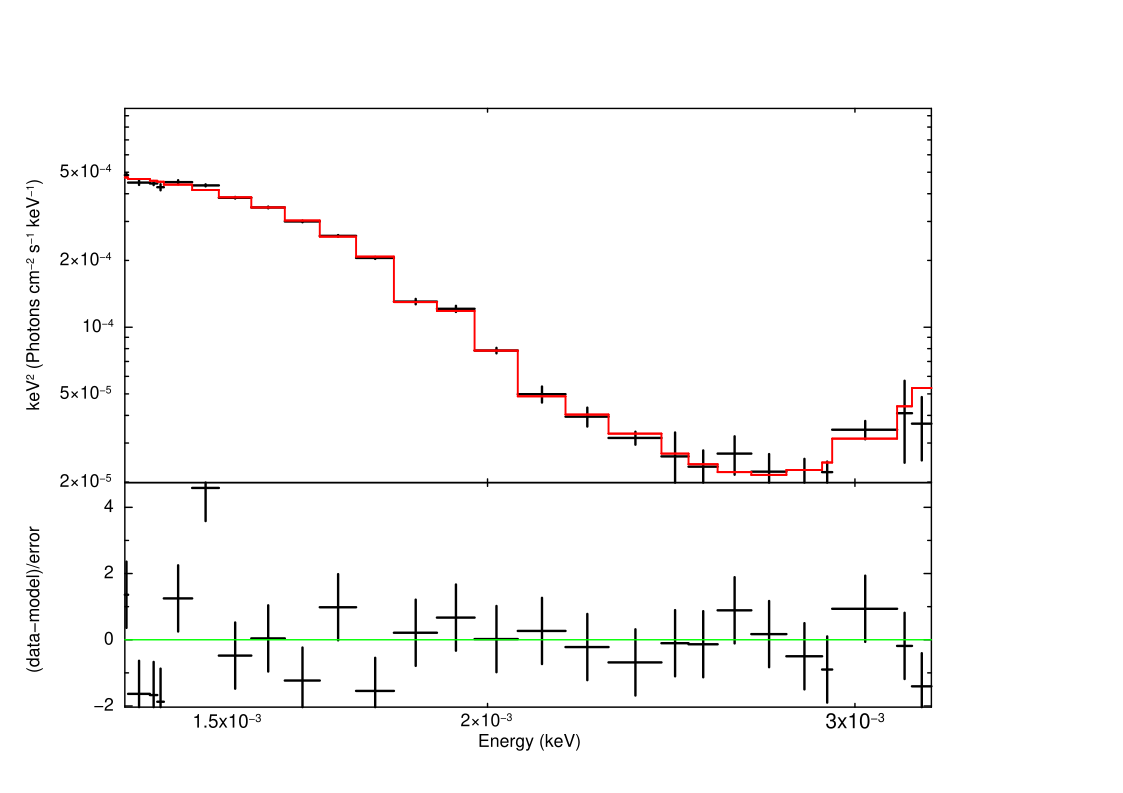}
 \caption{Energy flux $E F_{\rm E}$ of J1105 SDSS spectrum vs binned energy E. Top panel: Best-fit model (red line) to the data of J1105 (black crosses). Bottom panel: Residuals in terms of error. The reduced $\chi^2/dof$=1.64.}
\label{fig:best_fit_J1105_agnsed}
\end{figure}

The overall parameters of our best-fit model \texttt{redden*phabs*AGNSED*gabs(1)*gabs(2)} are given in Table~\ref{tab:agnsed_fitting_parameters}. The fixed values are clearly marked, and all errors were computed by XSPEC during the standard fitting process. The best-fit model versus data in the range of observational points is shown in Fig.~\ref{fig:best_fit_J1105_agnsed} together with the residuals, which are plotted in the bottom panel. Because we used dummy responses, the figure is made in physical units.
We did not consider photometric points taken from the SDSS or 2MASS surveys in our fitting procedure. Three arguments explain this approach. First, the SDSS optical photometric points lie within the errors on the SDSS spectrum, and they are very well represented by the SDSS spectrum. Including them would not improve our fit. Second, the photometric point lying at 580 \AA\ (in the rest frame of the quasar) does not have observational errors, and these could be very large and significantly distort the results. Finally, the 2MASS points show a significant effect of radiation from molecular-dust tori. Thus, the IR points require a good fit with a torus model. Most likely, there are two sources, and additional components in the best-fit model would need to be added. However, Fig. \ref{fig:sdss_J1105_plus_richards} shows that the number of photometric points and their spectral energy distribution coverage in the IR is too small. Modeling these points (we have only three) would therefore increase the errors of the final results. Profit will not be significant in the face of increasing parameter uncertainty.

\begin{figure}[ht]
 \centering
 \includegraphics[width=0.48\textwidth]{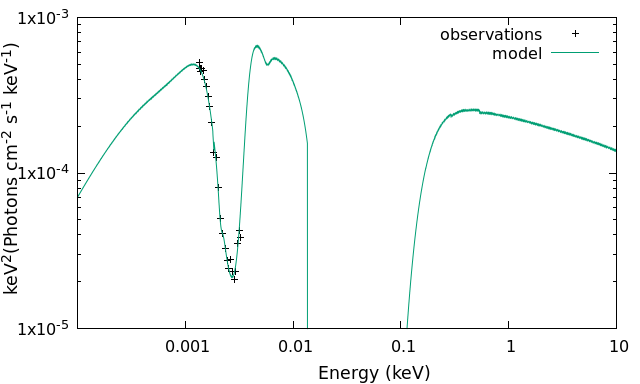}
\caption{Energy flux $E F_{\rm E}$ of the J1105 SDSS spectrum with the total best-fit model \texttt{redden*phabs*AGNSED*gabs(1)*gabs(2)} (green line) and the data of J1105 (crosses) in the broad energy range.}
\label{fig:obs_and_model}
\end{figure}

\begin{figure}[ht]
  \centering
  \includegraphics[width=0.48\textwidth]{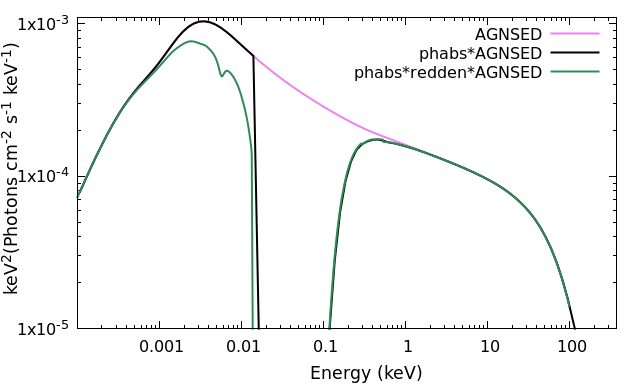}
  \caption{ Energy flux $E F_{\rm E}$ of the model components plotted individually in order to illustrate their  importance in comparison with the Gaussian functions. The pure \texttt{AGNSED} component is plotted by the pink line, the influence of Galactic absorption is shown by the black line, and the influence of reddening by the green  line. All models are plotted  based on the parameters taken from Table \ref{tab:agnsed_fitting_parameters}.}
\label{fig:agnsed}
\end{figure}

The unfolded best-fit model versus data for the broad range of energies is presented in Fig.~\ref{fig:obs_and_model} by a green line, where different absorption dips are visible. The data from the SDSS cover only one strong dip, so that in order to illustrate the origin of this dip, we present the decomposition into individual spectral components in Fig.~\ref{fig:agnsed}. The emission by the central nucleus is shown by the pink line. The strong dip between 0.01-0.1 keV is caused by photoelectric absorption in our Galaxy, and finally, reddening diminishes the energy between 0.001-0.01 keV, producing a small dip around 0.006 keV in the final spectrum, which is given by the green line. This plot clearly shows that the dip covered by the SDSS data of J1105 is purely modeled by two Gaussian functions. The reduced $\chi^2$ for the best fit is 1.64 for 12 degrees of freedom. We analyzed a similar approach with only one Gaussian function (i.e., N=1) to describe the deep absorption. However, the reduced $\chi^2$ value was very high - 84.020 for 8 degrees of freedom (when the model did not contain Gaussian components and all of the parameters were free). On the other hand, a further increase in the number of Gaussian functions does not significantly decrease the reduced $\chi^2$. Adding a third or fourth Gaussian function reduces the $\chi^2$/dof values by less than 10\%.

\begin{table}
    \centering
    \caption{Parameters of the best-fit model \texttt{redden*phabs*AGNSED*gabs(1)*gabs(2)} for J1105, where the number of absorbing Gaussian functions N=2. The hydrogen column density of the \texttt{phabs} model was calculated using NASA's HEASARC tool, and the reprocess parameter was  taken to be 1 (see Sec.~\ref{sec:xspec} for the discussion). The reduced $\chi^2/dof$=1.64..}
    \begin{tabular}{|c|c|c|}
    \hline
    Model &  Parameter & Fitted value \\
    component & and unit  &  and error\\
    \hline
    \hline\xrowht{10pt}
    \texttt{redden} & E(B-V) & 0.0087 (fixed) \\
    \hline\xrowht{10pt}
    \texttt{phabs} & $N_{\rm H}$ [cm$^{-2}$] & $8.63\times 10^{19}$ (fixed)\\
    \hline\xrowht{10pt}
    \multirow{14}{*}{\texttt{AGNSED}}  & \mbh\ [\Msun] & $3.52 \pm 0.01 \times 10^{9}$\\
    \cline{2-3}\xrowht{10pt}
    & \dotm & $0.274 \pm 0.001$\\
    \cline{2-3}\xrowht{10pt}
    & cos \incl & $0.854 \pm 0.015$ \\
    \cline{2-3}\xrowht{10pt}
    & $a_*$ & $0.32 \pm 0.04$ \\
    \cline{2-3}\xrowht{10pt}
    & $kT_{\rm e, hot}$ [keV] & 100 (fixed) \\
    \cline{2-3}\xrowht{10pt}
    & $kT_{\rm e, warm}$ [keV] & 0.50 (fixed) \\ 
    \cline{2-3}\xrowht{10pt}
    & $\Gamma_{\rm hot}$ & 2.20 (fixed) \\ 
    \cline{2-3}\xrowht{10pt}
    & $\Gamma_{\rm warm}$ & $2.86 \pm 0.22$ \\ 
    \cline{2-3}\xrowht{10pt}
    & $R_{\rm hot}$ [$r_g$] & $42.0 \pm  5.2$\\
    \cline{2-3}\xrowht{10pt}
    & $R_{\rm warm}$ [$r_g$] & $430.0^{+50.0}_{-11.0}$ \\ 
    \cline{2-3}\xrowht{10pt}
    & $R_{\rm out}$ [$r_g$] & 100000 (fixed) \\
    \cline{2-3}\xrowht{10pt}
    & $H_{\rm tmax}$ [$r_g$] &= $R_{\rm hot}$ \\
    \cline{2-3}\xrowht{10pt}
    & $D$ [Mpc] & 5075 (fixed) \\
    \cline{2-3}\xrowht{10pt}
    & $z_{\rm spec}$ & 1.929(fixed)\\
    \cline{2-3}\xrowht{10pt}
    & normalization & $0.409\pm0.148$ \\
    \hline\xrowht{10pt}
    \multirow{3}{*}{\texttt{gabs(1)}} & $E_1$ [keV] & $2.792 \pm 0.015 \times 10^{-3}$\\
    \cline{2-3}\xrowht{10pt}
    & $\sigma_1$ [keV] & $5.842\pm 0.069 \times 10^{-4}$\\
    \cline{2-3}\xrowht{10pt}
    & $\tau_1$ & $ 4.686 \pm 0.098 \times 10^{-3}$\\
    \hline\xrowht{10pt}
    \multirow{3}{*}{\texttt{gabs(2)}} & $E_2$ [keV] & $ 2.125 \pm 0.022 \times 10^{-3}$\\
    \cline{2-3}\xrowht{10pt}
    & $\sigma_2$ [keV] & $ 3.874 \pm 0.117 \times 10^{-4}$\\
    \cline{2-3}\xrowht{10pt}
    & $\tau_2$ & $ 7.025 \pm 0.367  \times 10^{-4}$ \\
    \hline
    \end{tabular}
    \label{tab:agnsed_fitting_parameters}
    \begin{quote}
    \end{quote}
    \end{table}
    
%===================================
\subsection{Central engine of J1105 }

We have no observations in X-ray or $\gamma$-ray, thus the obtained parameters that describe the hot corona and the warm skin are not reliably defined. 
Therefore, we fixed the parameters of the hot corona as $\Gamma_{\rm hot} = 2.20$, $kT_{\rm e, hot} = 100$ keV, and the temperature of the warm skin, $kT_{\rm e, warm} = 0.50$ keV, as typical values estimated for quasars  of similar SMBH mass $> 10^8$ \Msun (e.g., PG 1211+143, \citealt{Janiuk01}; PG 1048+213, \citealt{Done12}; Mrk 509, \citealt{Petrucci13}; RX J0439.6-5311, \citealt{Jin17}; PG 1115+407, \citealt{KubotaDone_2018}; Ark 120, \citealt{Porquet18}).

The photon index of the warm skin in J1105 from our fitting procedure is $\Gamma_{\rm warm}=2.86 \pm 0.22$. It is similar to the indices seen for PG 1115+407, where $\Gamma_{\rm warm}$ = 3.06 \citep{KubotaDone_2018} or for narrow-line type 1 QSO RX J0439.6-5311 with $\Gamma_{\rm warm}$ = 2.72 \citep{Jin17}.
 
The transition radius between the corona and the warm skin from our fitting is $R_{\rm hot}=42.0\pm5.2$ $r_{\rm g}$. By comparison,  for  BLS1 PG 1048+213 $R_{\rm hot} \sim 100\ r_{\rm g}$ \citep{Done12}, but lower values are found in other sources: $R_{\rm hot}=$ 15.1-41.8 $r_{\rm g}$ in PG 1448+273 \citep{Ding21} and $R_{\rm hot} \lesssim$ 10 $r_{\rm g}$ in PG 1211+143 (\citealt{Janiuk01}).

The radius at which the existence of the warm skin ends in J1105, $R_{\rm warm} = 430^{+50}_{-11}\ r_{\rm g}$. For individual AGNs, this value can be lower.  The beginning radius of the reflection component is estimated as $107^{+196}_{-52} \ r_{\rm g}$ for Ark 120 \citep{Porquet18}. This radius is $\sim 40 \ r_{\rm g}$ for Mrk~509 and $\sim151\ r_{\rm g}$ for the lower-mass NGC 5548 \citep{KubotaDone_2018}.
\cite{Czerny2003corona_highacc} set $R_{\rm warm}$ at about $600\ r_{\rm g}$ in the case of mean quasars. Thus, it is worth noting that the description of the corona/warm skin for J1105 is well established and comparable with literature studies.

%===============================================================================
\subsection{Absorbing clouds in J1105}
\label{sec:calc_abs}

Our result indicates that the deep absorption in J1105 could be explained by two Gaussian functions, which may describe absorption on two different clouds. All fitting parameters in the rest frame are given in Table~\ref{tab:observed_energy_rest_wave}. The minima of the Gaussian functions for J1105 in the rest frame are at $1516\pm10$ \AA, and $1992\pm20$ \AA\ with $\sigma$ $= 332\pm69$ and $375\pm 64$, receptively. The absorption described by the first \texttt{gabs(1)} dominates and has a strength of nearly 90\% of
the whole recess. The origin of this deep absorption in the J1105 is unclear. Here we focus on several possible scenarios depending on two assumed interpretations of the observed features.

In the beginning, we assume the presence of two systems. The proximity to the \CIV\ doublet ($\lambda$1548, 1551\AA) suggests that these lines might cause the observed absorption. When we assume that the global absorption in J1105 is caused by the extended clouds that absorb \CIV, then we are able to establish their shifts $z_{1, \mathrm{\CIV}} = - 0.0213$, and $z_{2, \mathrm{\CIV}} = + 0.2859$. The negative value of $z_i$ means that the absorbing material moves toward us. Corresponding velocities for \CIV\ absorbers computed with the use of Eq.~(\ref{eq:1}) are given in the fifth column of Table~\ref{tab:observed_energy_rest_wave}.

With the use of the \CIV\ line, the first cloud may indicate the presence of intrinsic absorption with an outflowing velocity of about $6330$ \kms. The second redshifted cloud would rather be associated gas falling toward the quasar nucleus with a speed of $\sim 73887$ \kms, most probably connected with the FRADO proposed by \cite{Czerny2018Frado}. This outflow participates in the formation of the BLR and finally falls onto the accretion disk. Nevertheless, the speed of the FRADO is about 1000 \kms\ \cite{Naddaf_Czerny2021Frado, Czerny_Hryniewicz2011_failed_wind}, one order of magnitude lower than our second component. The speed of the FRADO modeled by these authors is only the lower limit, however, because the acceleration of the material by the line-driving mechanism is not taken into account in the computations, even when the lines are commonly observed in the quasar spectra. The FRADO can be a possible interpretation of the observed absorption in J1105. Therefore, further calculations should be made to solve it.

\begin{table*}
    \centering
    \caption{Parameters of two Gaussian functions  \texttt{gabs(N)} fitted to our data 
    at the rest frame ($z_{spec} = 1.929$) given in keV and \AA\ in the second and third column, respectively. The equivalent width was computed numerically on the unconvolved fitted model and is given in the fourth column. The velocity and ionic column density for the \CIV\  absorber is shown in the fifth and sixth column and for the \MgII\  absorber in the seventh and eighth column. A negative value of the velocity indicates that matter is outflowing toward the observer.}
    \begin{tabular}{c|c|c|c|c|c|c|c}
    \multirow{2}{*}{N} & Rest  & Rest  & $EW$  & \CIV\  ($\lambda$1548, 1551 \AA)  &  $\log(N_{\CIV\ })$ & \MgII\   ($\lambda$2796, 2803 \AA) &   $\log(N_{\MgII\ })$  \\
    &  $\times 10^{-3}$ [keV] & [\AA] & [\AA] & $V_{\rm out} $ [\kms] &  [cm$^{-2}$] & $V_{\rm out} $ [\kms] & [cm$^{-2}$] \\
    \hline
    (1) & (2) & (3) & (4) & (5) & (6) & (7) & (8) \\
    \hline
     \hline
    $ E_1$ & $ 8.178 \pm 0.054   $ & $1516\pm10$ & \multirow{2}{*}{$714^{+145}_{-147}$} & \multirow{2}{*}{$-6330^{+1698}_{-1710}$} & \multirow{2}{*}{$17.027^{+0.080}_{-0.010}$} & \multirow{2}{*}{$-108135^{+555}_{-615}$} & \multirow{2}{*}{$16.003^{+0.080}_{-0.010}$} \\
    \cline{1-3}
    $\sigma_1$ &  $ 1.711\pm 0.020 $ & $332\pm69$ & &  & &  & \\
   \hline
      $E_2$  & $6.224\pm0.062  $ & $1992\pm20$ & \multirow{2}{*}{$124^{+20}_{-21}$}  & \multirow{2}{*}{$73887^{+2827}_{-2871}$} &  \multirow{2}{*}{$16.266^{+0.065}_{-0.080}$} & \multirow{2}{*}{$-74447^{+1543}_{-1577}$} &  \multirow{2}{*}{$15.242^{+0.065}_{-0.080}$} \\
    \cline{1-3}
    $\sigma_2$  & $1.135\pm 0.033  $ &  $375\pm 64$ & &  & & & 
    \end{tabular}  
    \label{tab:observed_energy_rest_wave}
\end{table*}

When we assume that the absorption in J1105 is caused by \CIV, the BI computed based on the blueshifted lines alone is equal to  20300 \kms. This value is higher than the BI of the strongest absorbers detected so far \citep{Paris2018}, but not impossible because at least 13 sources out of 21877 reported by these authors have a BI in the range of 20300-23758 \kms.

\begin{figure}
   \centering
    \includegraphics[width=0.4\textwidth]{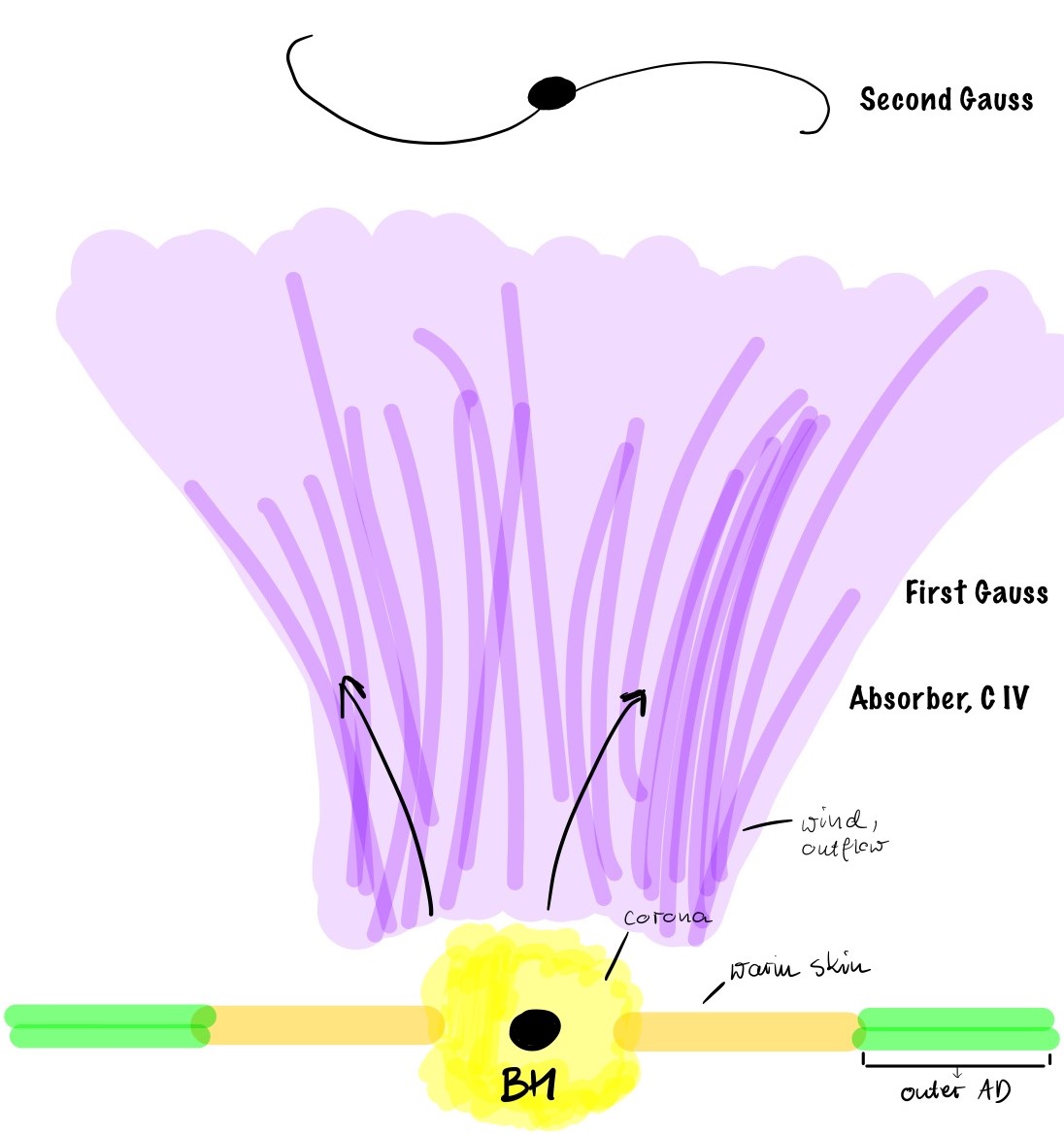}
 \caption{Possible scenario of absorption in J1105 interpreted with the \CIV\ line. Absorbing material is outflowing from the system. It partially turns back and builds up the FRADO, and partially leaves the quasar. The outflowing material may feed the local (to the quasar) normal or dwarf galaxy and cause additional absorption.}
 \label{fig:J1105_wind}
\end{figure}

On the basis of our study, it is difficult to estimate the location of the absorbing material. We can only propose the scenario that the first Gaussian arises during absorption of radiation from the central engine by the wind that gives a rise to BAL and falls back onto the accretion disk as FRADO, while the second Gaussian can arise as absorption on a more distant gas cloud or dwarf galaxy, but still at the redshift of the observed quasar. We present the illustration of this scenario in Fig.~\ref{fig:J1105_wind}.

\begin{figure}
  \centering
    \includegraphics[width=0.4\textwidth]{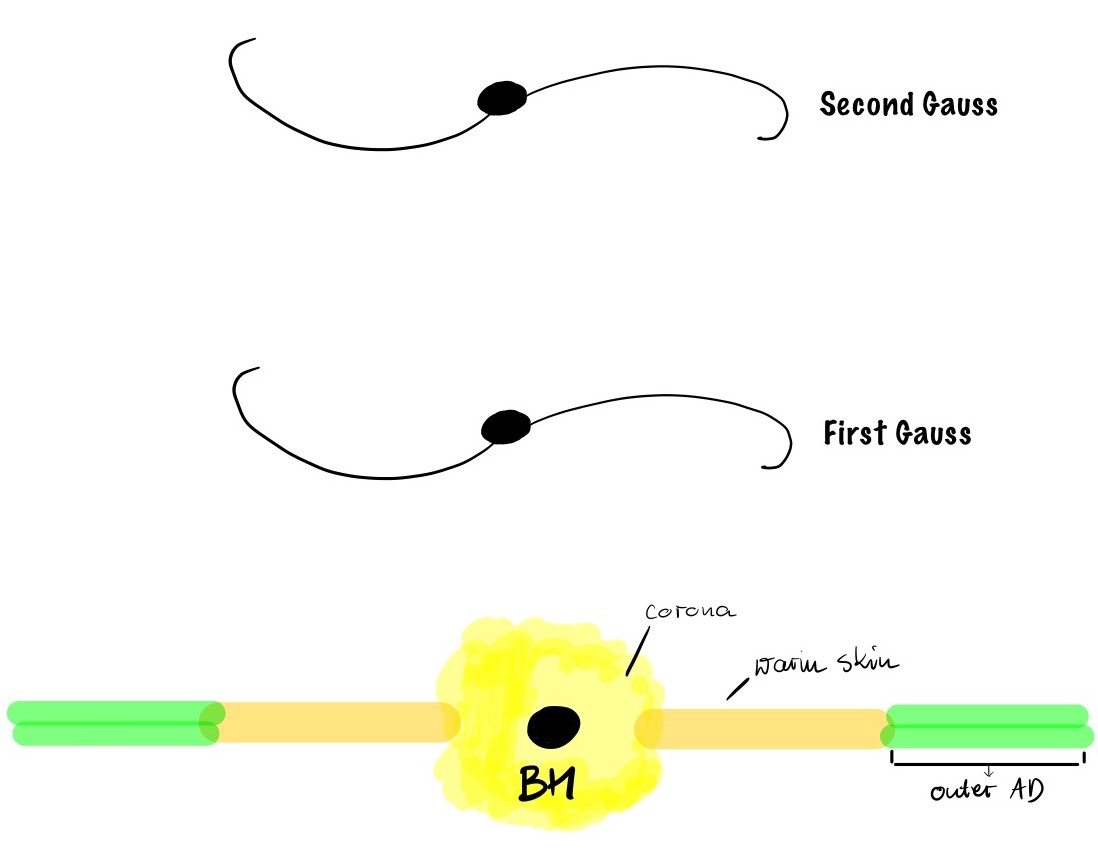}
\caption{Possible scenario of absorption in J1105 interpreted with the \MgII\ line. The outflowing material may feed the local (to the quasar) normal or dwarf galaxies and cause absorption. An explanation of the nature of J1105 is that it consists of two normal or dwarf galaxies.}
\label{fig:J1105_2galaxies}
\end{figure}

Next, we assumed that both systems exhibit absorption in \MgII\ ($\lambda$2796, 2803 \AA), which is the second readily absorbed doublet. Then their  $z_{1, \mathrm{\MgII}} = - 0.4585$, and $z_{2, \mathrm{\MgII}} = - 0.2885$ for the first and second absorber, respectively. The corresponding velocities are listed in the seventh column of Table~\ref{tab:observed_energy_rest_wave}. If the deep absorption in J1105 is caused by the \MgII\ alone, then the situation is more clear. According to the high velocities, two absorbing clouds would be related to intervening systems that lie along the line of sight to the observer. These clouds may be associated with dwarf galaxies detached from the quasar and located on the way toward the observer, as illustrated in Fig.~\ref{fig:J1105_2galaxies}.

We have developed an algorithm to numerically compute the $EW$ of the absorbing features  we fit in the unfolded best-fit model. The results for two Gaussian functions are presented in the fourth column of Table~\ref{tab:observed_energy_rest_wave}. Next, the $EW$ values were used to estimate the ionic column densities of the absorbing systems, following Eq.~(\ref{eq:2}). The ion column densities are high, as expected for the deep absorption in J1105. This would imply that the total column density of the absorbing gas is high, suggesting that the source may be covered by gas with a high optical depth that should be seen in the X-ray spectrum as Compton-thick emission. X-ray detection is needed to solve this issue in future studies.

%=================
\subsection{Black hole mass in J1105} 

The fitting procedure performed by us in this paper allowed for an indirect estimation of the SMBH mass and accretion rate because both quantities are free parameters in the \texttt{AGNSED} model implemented in XSPEC. We obtained $\log M_{\rm BH} = 9.547 \pm 0.002$ and $\log \dot{m} = -0.562 \pm 0.002$. These two values lie within a few $\sigma$ from the values $\log M_{\rm BH} = 11.205$ and $\log \dot{m} = -4.713$ reported by \cite{Shen2011qsocatalog}. The authors pointed out that the two estimated values were obtained using the relation between the BH mass or accretion rate and the FWHM and luminosity of a given emission line ($L_{\lambda}$), and they are to be treated with caution. 

Recently, it was reported by \cite{Mejia18N} in the case of type 1 AGN and by \cite{Marculewicz2020} for WLQ that the $M_{\rm BH}$ and \dotm\ values in the relation mentioned above are biased. The reason is the poorly known geometry and dynamic of the BLR clouds that produce the emission lines. In addition, strong absorption may influence the detection of emission line profiles, and hence their velocities. 

The observed narrow emission lines in J1105 (see Fig.~\ref{fig:sdss_J1105_plus_richards})
most likely originate in the narrow line region (e.g., FWHM(\CIV) $= 782 \pm 279$ \kms\ ; \citealt{Shen2011qsocatalog}). Therefore, the emission lines from the BLR were never observed, and we can only assume that they may reach velocities comparable to our winds described in the section above. With this crude assumption, we may suspect that BLR lines may reach velocities of $\sim 10^5$~\kms . According to the formulas used in \citet[][see Fig.2 for illustration]{Mejia18N}, this means that the real BH mass is two orders of magnitude lower than the so-called virial mass reported by \cite{Shen2011qsocatalog}. Follwing \citet{Mejia18N}, we adopted a virial factor $f$, that is, the ratio of $M_{\rm BH}$ to the SE virial mass estimator, that is inversely proportional to the FWHM for all lines, and it is 1 for FWHM$=3200 \pm 800$~\kms \ in case of \MgII.

We obtain this BH mass, two orders of magnitude lower than the virial mass, from our continuum-fitting method. This result only fits in the case of our fastest wind reported in Table~\ref{tab:observed_energy_rest_wave}, and it clearly indicates the correctness of the work made by authors who aimed to find the proper relation between the SE virial mass estimation and $M_{\rm BH}$. Even though our method does not use the FWHM (line) to calculate \mbh\ or \dotm, the total model depends on many parameters, and our BH mass estimation is just an exemplary exercise and may deliver different results, while X-ray data will be available for J1105 in the near future.

%========================================== 

\subsection{Mass outflow rates}

The outflow velocities obtained by us have rather high values, therefore they may produce a huge mass outflow from the quasar into the intergalactic medium. Following the commonly used procedure, we computed overall mass outflow rates for J1105 based on the measured bolometric luminosities, given in Sec.~\ref{source}, and the velocities reported in Table~\ref{tab:observed_energy_rest_wave}. We describe our results in Table~\ref{outfow:rates}. We note that our continuum-fitting method allows us to compute the total luminosity from a measured accretion rate, which in case of a typical accretion efficiency between 0.06-0.08 is $\sim1.278 \times 10^{47}$ \ergs, that is, an order of magnitude higher than the bolometric luminosity measured from a monochromatic source luminosity at 3000\AA\ (see Sec.~\ref{source}).

Nevertheless, we lack information about the radiation that is released when accretion interacts with wind, therefore we should treat this value with caution. We adopted the value of $L_{\rm bol}$ calculated from the luminosity at 3000\AA\ as more realistic, but we recall that the bolometric luminosity taken from the value of the accretion rate produces a mass outflow of about 106.81~\Msyr\ for the stronger \CIV(1) absorber (visible at 1516 \AA). 
\cite{Tombesi2012} observed 42 local radio-quiet AGNs and constrained the mass outflow rates between 0.01 and 1 \Msyr, which is slightly lower than those of J1105. The first indication is that the difference may come from the redshift, that is, the local radio-quiet sample versus z = 1.929 of J1105. This conclusion can be confirmed by the recent results obtained from 14 QSOs in the redshift range of 1.41--3.91, which indicates that outflows with higher mass outflow rates reached even $\dot M_{\rm out} = 101 ^{+59}_{-53}$~\Msyr\ in case of APM 08279+5255, classified as BALQSO with $\log M_{\rm BH} = 10.0$~[\Msun] \citep[][Table 9]{Chartas2021_UFO}. All 14 reported quasars have measured mass outflow rates 
in the range from $0.21-101$~\Msyr, which is consistent with our result. This indicates similarities with J1105 regarding the redshift and value of the outflow.

\begin{table}
    \centering
    \caption{Mass outflow rates computed for two absorbers given by N, interpreted by the two lines. The rates depend on the two values of bolometric luminosities calculated in Sec.~\ref{source}. 
    A negative value means that the mass is outflowing toward observer.}
  \begin{tabular}{c|c|c|c}  
  Absorber & N  & $\log L_{\rm bol}$ [\ergs] & $\dot M_{\rm out}$ 
  [\Msyr] \\
  \hline
  \hline\xrowht{12pt}
  \multirow{4}{*}{\CIV\ } & \multirow{2}{*}{1} & 46.106 & $-10.659^{+3.945}_{-2.255}$ \\
   \cline{3-4}\xrowht{12pt}
    & & 46.225 & $-14.019^{+5.188}_{-2.965}$ \\
    \cline{2-4}\xrowht{12pt}
   & \multirow{2}{*}{2} & 46.106 & $0.913^{+0.369}_{-0.336}$ \\
    \cline{3-4}\xrowht{12pt}
   & & 46.225 & $1.201^{+0.048}_{-0.044}$ \\
   \hline\xrowht{12pt}
   \multirow{4}{*}{\MgII\ } & \multirow{2}{*}{1} & 46.106 & $-0.624^{+0.004}_{-0.003}$ \\
  \cline{3-4}\xrowht{12pt}
   & & 46.225 & $-0.820^{+0.005}_{-0.004}$ \\
   \cline{2-4}\xrowht{12pt}
   & \multirow{2}{*}{2} & 46.106 & $-0.906^{+0.019}_{-0.018}$ \\
   \cline{3-4}\xrowht{12pt}
   & & 46.225 & $-1.192^{+0.026}_{-0.024}$ \\
   \hline 
  \end{tabular}
  \label{outfow:rates}
\end{table}

The typical value of the mass outflow rate for sources accreting below or close to the Eddington limit is $\dot M_{\rm out} \geq$ 5-10\% $\dot M_{\rm acc}$ (where $\dot M_{\rm acc}$  is the mass accretion rate in \Msyr) for both UFOs and non-UFOs \citep{Tombesi2012}. Taking into account our fitted value of the accretion rate of J1105 ($\dot m = 0.274$, which corresponds to $\dot M_{\rm acc} = 36.416 \pm 0.133 $~\Msyr), we achieved a huge mechanical power $\dot M_{\rm out} \geq$ 1.7-38.5\% $\dot M_{\rm acc}$  in case of three outflows (\CIV(1), \MgII(1), and \MgII(2) from Table~\ref{outfow:rates}) observed in our quasar. This mechanical power may cause a significant feedback impact on the surrounding medium. We conclude here that highly absorbed quasars may produce massive outflows, which may reach velocities comparable to UFOs. Nevertheless, observations of J1105 in the X-ray domain are needed to confirm this statement. 

Massive outflows were also detected in the past on the basis of observations of neutral gas in \HI\ absorption, indicating  $\dot M_{\rm out} \sim 50$~\Msyr\ \citep{Morganti2005} and in starburst-driven super-winds in ultraluminous IR galaxies (ULIRG), which in general show a mass outflow range between 
$\sim 10-1200$~\Msyr\ \citep{Rupke2005,Netzer13,Tombesi2016}. The origin of these outflows was explained based on the  fast winds, which are strong evidence of the interaction between radio jets and the surrounding interstellar medium. In the future development of J1105 with new observations in all energy domains, we can verify the presence of a jet-driven outflow.

%==============================================================================
\section{Discussion and conclusions}
\label{sec:5}

We analyzed the absorbed optical/UV spectrum of the SDSS~J110511.15+530806.5 quasar, which was not extensively observed in the past. Our analysis was made assuming that the central engine consists of an accretion disk and an eventual warm corona, but we lacked points in the X-ray domain. This was done to explain the deep absorption in the considered quasar when the continuum model has to be fitted first. Our approach assumed that this deep absorption can be explained by two broad Gaussian functions in addition to the \texttt{AGNSED} model that is available in the XSPEC fitting package. First, we checked that this deep absorption, which is not observed in other quasars, cannot be explained by different extinction laws for the Galaxy dust and grains.

In order to explain the observed absorption, we modeled the intrinsic emission from the source nucleus. In the case of J1105, the intrinsic radiation of the central engine was analyzed as the sum of emission from an accretion disk, warm skin, and hot corona components with the standard XSPEC model. The fit parameters that describe the corona and warm skin in the case of our quasar should be treated with caution, however, because even though they are comparable with the literature studies of similar objects, we lack X-ray observations of J1105. The overall spectral shape of J1105 in  X-ray domain is needed to place final constraints on its central engine parameters. 

In the next step of our analysis, the deep absorption was represented by two Gaussian functions that reflect two absorbing systems presented on the way toward an observer. Further conclusions depend on the ion by which we identify the two absorbers. With the use of the \CIV\, line, the first and dominant Gaussian function represents an outflow toward us with a speed of about 6330 \kms. This wind may be associated with intrinsic absorption that is typically observed in QSOs and is caused by a distant wind originating from an accretion disk or disk and jet interaction. The speed of the second absorber is high, about 74000 \kms, and speaks for gas moving toward the center of the quasars. Most probably, this is connected with the FRADO proposed by \cite{Czerny2018Frado}. This outflow participates in the formation of the BLR and finally falls onto the accretion disk. Nevertheless, the speed of the FRADO is about 1000 \kms\ \citep{Naddaf_Czerny2021Frado, Czerny_Hryniewicz2011_failed_wind}, one order of magnitude lower than in the case of the velocity found in this paper. On the other hand, the speed of the FRADO modeled by the above authors is only a lower limit because the acceleration of the absorbing gas by the line-driving mechanism was not taken into account in their computations \citep[][private com.]{Naddaf_Czerny2021Frado}. Simulations show that line-driven winds may accelerate gas to huge velocities \citep{proga2000}, and our measurement may be physically possible. Further theoretical achievement in estimating the FRADO velocities that includes line-driven acceleration may solve this issue in the near future. 

Based on the \MgII\ line, both velocities point out the intervening systems with the quasar gas moving toward us with speeds of about 108000 and 74000 \kms\  for the first and second absorber, respectively. Fast winds like this produce a high mass outflow rate that in the case of our fastest wind reaches 38.5\% of the disk accretion rate derived from the fit. This massive outflow may cause a significant feedback impact on the surrounding medium. We conclude here that highly absorbed quasars may produce massive outflows, which may reach velocities comparable to those of UFOs. \cite{Tombesi2013} examined a Seyfert 1 sample with velocities in the range $10^4 - 10^5 \mkms$, exactly the values obtained for J1105 in the case of an \MgII\ absorber. This might suggest that the UFO concept is a plausible explanation of the abnormal properties of the SDSS spectrum of J1105. 

The massiveness of the outflow detected in this paper manifests itself at mass outflow rates up to 
$M_{\rm out} = 14.019$~\Msyr\ computed on the basis of the bolometric luminosity measurements for J1105, and even up to $106.814$~\Msyr\ for the bolometric luminosity computed from the accretion rate obtained from the continuum fitting analysis. A high mass outflow rate is seen in more ULIRGs, which show outflows rates ranging from 10 to 1000 \Msyr\ \citep[e.g.,][]{Rupke2005}. Additionally, \cite{Netzer13} indicated that some of the ULIRGs have outflow velocities exceeding 1000 \kms\ and mass outflow rates $\sim$ 1200~\Msyr. The origin of these winds is still unknown, but in the case of ULIRG sources, it is often interpreted as an interaction of the jet with the surrounding interstellar medium.

The ion column densities in the two absorbers are high, as expected for the deep absorption in J1105. For example, \cite{Saez2021_NH} indicated that the values of the logarithms of the column density of CIV in BAL QSO PG 2112+059 are in the range 15.38-15.81, and they varied over 13 years. Additionally, when they assumed that the absorber was seen partially and the covering factor of the absorber was $\simeq$ 60\%, then the mean log $N(\CIV)$ of the absorber was even higher and equal to 16.12. We did not consider any partial covering in our analysis so far. This point will be included when better spectra are available from future monitoring of the source. For $N_{\MgII}$ , the ion column density is also high in the J1105 quasar. However, $\log (N_{\MgII})$ seen in the system B of HS 1603+3820 quasar is $15.096\pm0.159$ \citep{Dobrzycki2007} and is comparable within the 1$\sigma$ error with the value obtained in this paper for the J1105 quasar.

When we take into account that the absorption is caused by the \CIV\ doublet, the BI is equal to 20300 \kms for the quasar considered in this paper. This value is as high as the BI seen in the strongest BAL QSOs. In the SDSS quasar catalog DR14 \citep{Paris2018}, only 0.06\% (13/21877) of the BAL quasars have this index lying between 20300 and 23758 \kms. Future observations of the absorbing features in J1105 may prove our hypothesis that the quasar has an extremely strong BAL.

The continuum-fitting method allowed us to estimate the BH mass and accretion rate in J1105. We obtained $\log M_{\rm BH} = 9.547 \pm 0.002$ and $\log \dot{m} = -0.562 \pm 0.002$, values that differ from previous estimations. The BH mass is two orders of magnitude lower than previously reported \citep[$\log M_{\rm BH} = 11.205$]{Shen2011qsocatalog}. 
Moreover, the accretion rate obtained by us shifts the quasar toward bright objects that radiate away 20\% of their accretion power. These results need to be confirmed by future studies with X-ray data of J1105. 

The BH mass of J1105 we found needs special attention. We showed that the fasted wind velocity detected by us, that is, $\sim 10^5$~\kms, can be attributed to the BLR, therefore the virial factor examined by \citet{Mejia18N} is about $10^{-2}$. This virial index means that the real BH mass is exactly two orders of magnitude lower than the virial mass found with RM and SE measurements. This is what we obtained here, and this result confirms the hypothesis that BH mass estimations based on lines are affected by the presence of wind and should be corrected according to formulas given 
by \citet{Mejia18N} in the case of normal AGNs and by \citet{Marculewicz2020} in case of WLQ. This result is also intriguing from another point of view. It indicates that the strong absorption, however, is due to the fast wind and not to the presence of a dwarf galaxy along the line of sight of the J1105 quasar.

We take the following message for the future from our studies. The explanation of the origin of the absorption may lie in new X-ray, optical, UV, infrared, and radio observations, which we are planning to conduct.

Particularly the Hubble Space Telescope or James Webb Space Telescope would be very helpful to deliver the best-resolution data from distant sources such as the quasar presented in this paper. With the use of new data, it would be important to determine whether the outflow of J1105 is caused similarly to the \HI\ outflow by the mechanism that indicates the interaction between radio jets and the surrounding interstellar medium. It is necessary to determine how the outflow mass rate in J1105 relates to the global paradigm of deep unusual absorption with high-velocity winds. In the future development of J1105 and with new observations, we will be able to determine whether X-ray and/or radio observations contain an indication of the influence of a jet-driven outflow. It would be possible to determine the properties of the central engine and the BH mass with the correct virial factor in highly absorbed quasars if we collected a sample of these objects.

\begin{acknowledgements}  
We would like to thank the anonymous referee for useful comments that improved the clarity of the paper. We also thank Bożena Czerny, Krzysztof Hryniewicz, and Raj Prince for helpful feedback and discussions. AR was partially funded by National Science Center grant Nr. 2021/41/B/ST9/04110. This publication makes use of data from the Sloan Digital Sky Survey (SDSS) and the NASA/IPAC Extragalactic Database (NED).
\end{acknowledgements}

\bibliography{aa}
\end{document}